\input{aipcheck}

\documentclass[
    ,final            % use final for the camera ready runs
  ]
  {aipproc}

\layoutstyle{8x11single}

\begin{document}

\title{DN interaction from  the J\"ulich meson-exchange model}


\author{L Tolos}{
  address={Institut de Ci\`encies de l' Espai (IEEC/CSIC), Campus Universitat 
Aut\`onoma de Barcelona, Facultat de Ci\`encies, Torre C5, E-08193 Bellaterra 
(Barcelona), Spain}
}

\author{J. Haidenbauer }{
  address={Institute for Advanced Simulation,
Forschungszentrum J\"ulich, D-52425 J\"ulich, Germany}
}

\author{G. Krein}{
address={Instituto de F\'{\i}sica Te\'{o}rica, Universidade Estadual
Paulista,
Rua Dr. Bento Teobaldo Ferraz, 271 -  01140-070 S\~{a}o Paulo, SP, Brazil 
}
}

%\author{Ulf-G. Mei{\ss}ner}{
 % address={Institute for Advanced Simulation,
%Forschungszentrum J\"ulich, D-52425 J\"ulich, Germany},
%altaddress={Helmholtz-Institut f\"ur Strahlen- und Kernphysik (Theorie) and Bethe Center for Theoretical Physics, Universit\"at Bonn, Nu\ss allee 14-16, D-53115 Bonn, Germany }
%}

\begin{abstract}
The $D N$ interaction is studied in close analogy to the meson-exchange $\bar KN$ 
potential of the J\"ulich group using SU(4) symmetry constraints.  The model generates the $\Lambda_c$(2595) resonance dynamically as a $DN$ quasi-bound state.
Results for $DN$ scattering lengths and cross sections are presented
and compared with predictions based on the Weinberg-Tomozawa term.
Some features of the $\Lambda_c$(2595) resonance are also discussed emphasizing
the role of the near-by $\pi\Sigma_c$ threshold. 
 \end{abstract}

\maketitle

%%%%%%%%%%%%%%%%%%%%%%%%%%%%%%%%%%%%%%%%%%%%
%% MAINMATTER
%%%%%%%%%%%%%%%%%%%%%%%%%%%%%%%%%%%%%%%%%%%%

\section{Introduction}

The study of the interaction of open charm $D$-mesons with nucleons 
is challenging for several reasons. From the experimental side, reliable models
are crucial for guiding planned experiments by the 
$\bar{\rm P}$ANDA and CBM experiments of the future FAIR facility at Darmstadt~\cite{FAIR}. 
From the theoretical point of view, the physics motivations are several.
Amongst the most exciting ones is the possibility of studying 
chiral symmetry in matter. Also, studies of $J/\psi$ dissociation 
in matter \cite{Matsui:1986dk} require a good knowledge of the 
interaction of $D$-mesons with ordinary hadrons. Another exciting 
perspective is the possibility of the formation of $D$-mesic 
nuclei~\cite{Tsushima:1998ru,GarciaRecio:2010vt} 
and of exotic nuclear bound states like $J/\psi$ binding to 
nuclei~\cite{Brodsky:1989jd,Luke92,Tsu11}. 

To this end, coupled-channel meson-baryon models in the charm sector have been developed \cite{Tolos04,Lutz04,Hofmann05,Mizutani06,Lutz06,GarciaRecio09,JimenezTejero}. In those approaches the strong attraction is provided by vector-meson 
exchange \cite{Hofmann05,JimenezTejero},  by the Weinberg-Tomazawa (WT) term \cite{Lutz04,Tolos04,Mizutani06,Lutz06,Tolos08}, or by an extension of the WT interaction to an SU(8) 
spin-flavor scheme \cite{GarciaRecio09,Tolos10}.  All of them obtained dynamically the $\Lambda_c$(2595) resonance. This resonance was reported
by the CLEO collaboration \cite{CLEO1} and subsequently confirmed by several
other experiments \cite{E6872,ARGUS2,CLEO2}. 

In this paper, the $DN$ interaction is derived in close analogy to the meson-exchange 
$\bar K N$ model of the J\"ulich group \cite{MG}. Results for  $DN$ scattering observables are obtained and compared to the outcome of the leading-order SU(4) WT contact term \cite{Mizutani06,Tolos08} and to a SU(8) WT scheme \cite{GarciaRecio09,Tolos10}. We also analyze 
 the $\Lambda_c$(2595) resonance and focus on the consequences of the fact that this resonance coincides practically with the $\pi\Sigma_c$ threshold.

  \section{DN model in the meson-exchange framework}
  
The $DN$ interaction is derived in close analogy to the meson-exchange 
$\bar K N$ model of the J\"ulich group \cite{MG}, using as a working hypothesis SU(4) symmetry constraints, and by exploiting also the close connection between the $DN$ and $\bar DN$ systems due to G-parity conservation. More specifically, we use the latter constraint to fix the contributions to the direct $DN$ interaction potential while the former one provides the transitions to channels that can couple to the $DN$ system. The main ingredients of the $DN\to DN$ interaction are provided by vector meson ($\rho$, $\omega$) exchange and higher-order box diagrams involving ${D}^*N$,
$D \Delta$, and ${D}^*\Delta$ intermediate states.  As far as the coupling to other channels is concerned, we follow here Ref.~\cite{MG} and take into account only the channels
$\pi\Lambda_c(2285)$ and $\pi\Sigma_c(2455)$. Furthermore, we restrict ourselves
to vector-meson exchange and we do not consider any higher-order diagrams in
those channels. Pole diagrams due to the $\Lambda_c(2285)$ and $\Sigma_c(2455)$
intermediate states are, however, consistently included in all channels. We refer  the reader to Ref.~\cite{haid11} for details. The resulting interaction potential ${\cal V}_{ij}$ 
($i,j=$ $DN$, $\pi\Lambda_c$, $\pi\Sigma_c$) is then used to calculated
the corresponding reaction amplitudes ${\cal T}_{ij}$ by solving a coupled-channel Lippmann-Schwinger-type scattering equation:
\begin{eqnarray}
{\cal T}_{ij} = {\cal V}_{ij} + \sum_k {\cal V}_{ik} {\cal G}^0_k {\cal T}_{kj} \ ,
\label{LSE}
\end{eqnarray}
from which we calculate the observables in the standard way \cite{Juel1}.

\section{DN scattering observables and the $\Lambda_c(2595)$ resonance}

\begin{figure}[t]
\begin{minipage}{0.33\textwidth}
\includegraphics[height=6.5cm,angle=-90]{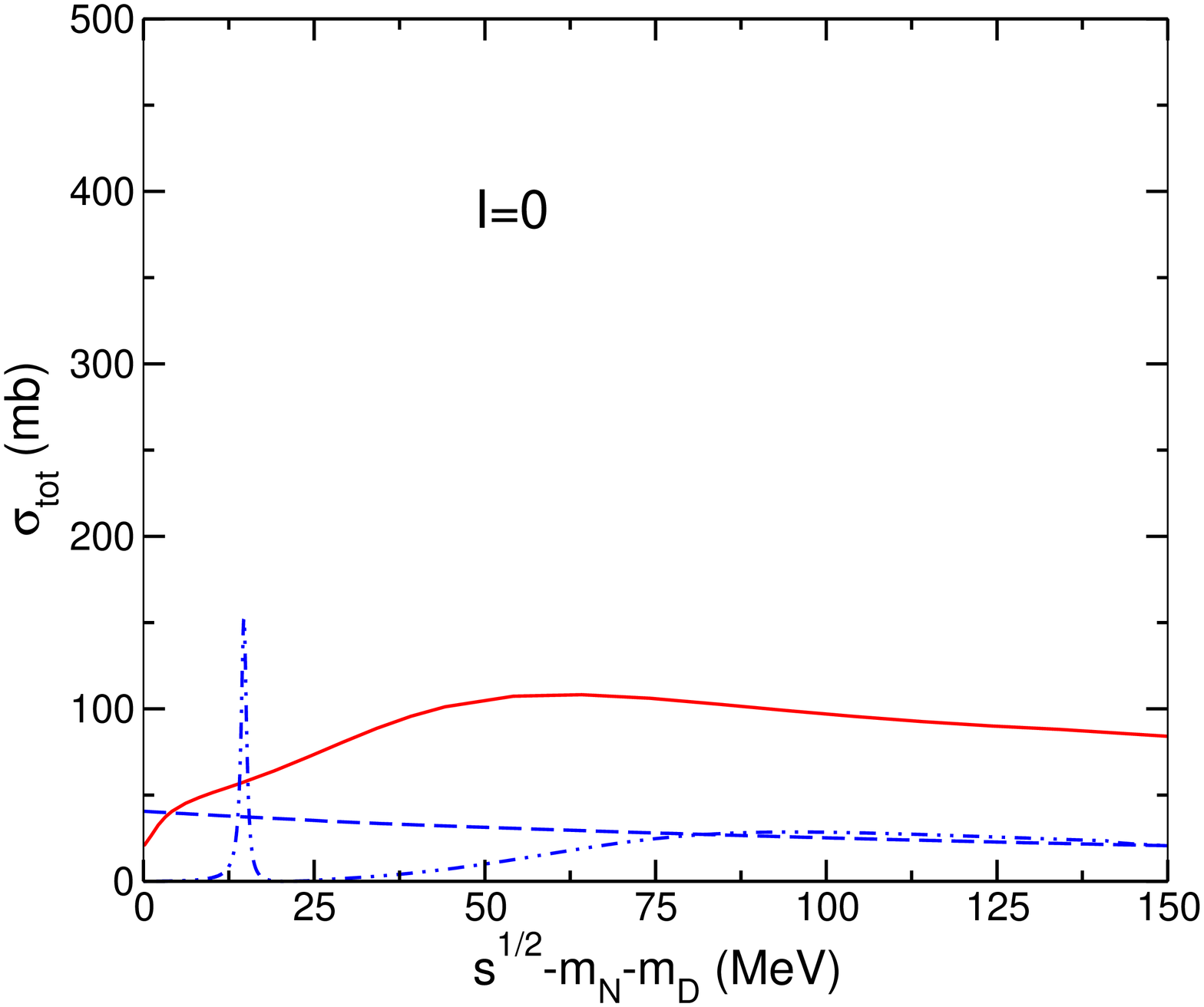}
\end{minipage}
\begin{minipage}{0.33\textwidth}
\includegraphics[height=6.5cm,angle=-90]{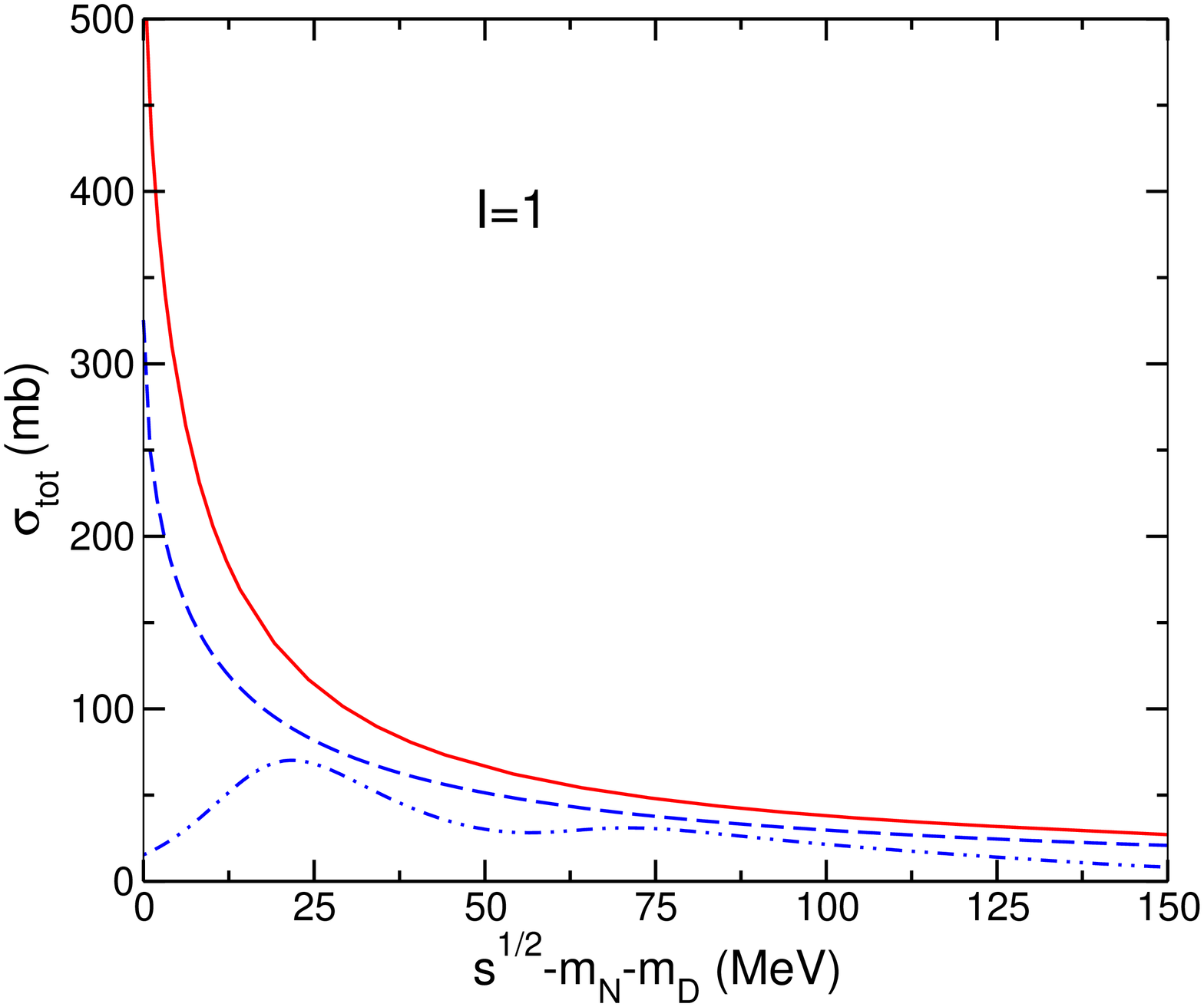}
\end{minipage}
\begin{minipage}{0.33\textwidth}
\includegraphics[height=8cm]{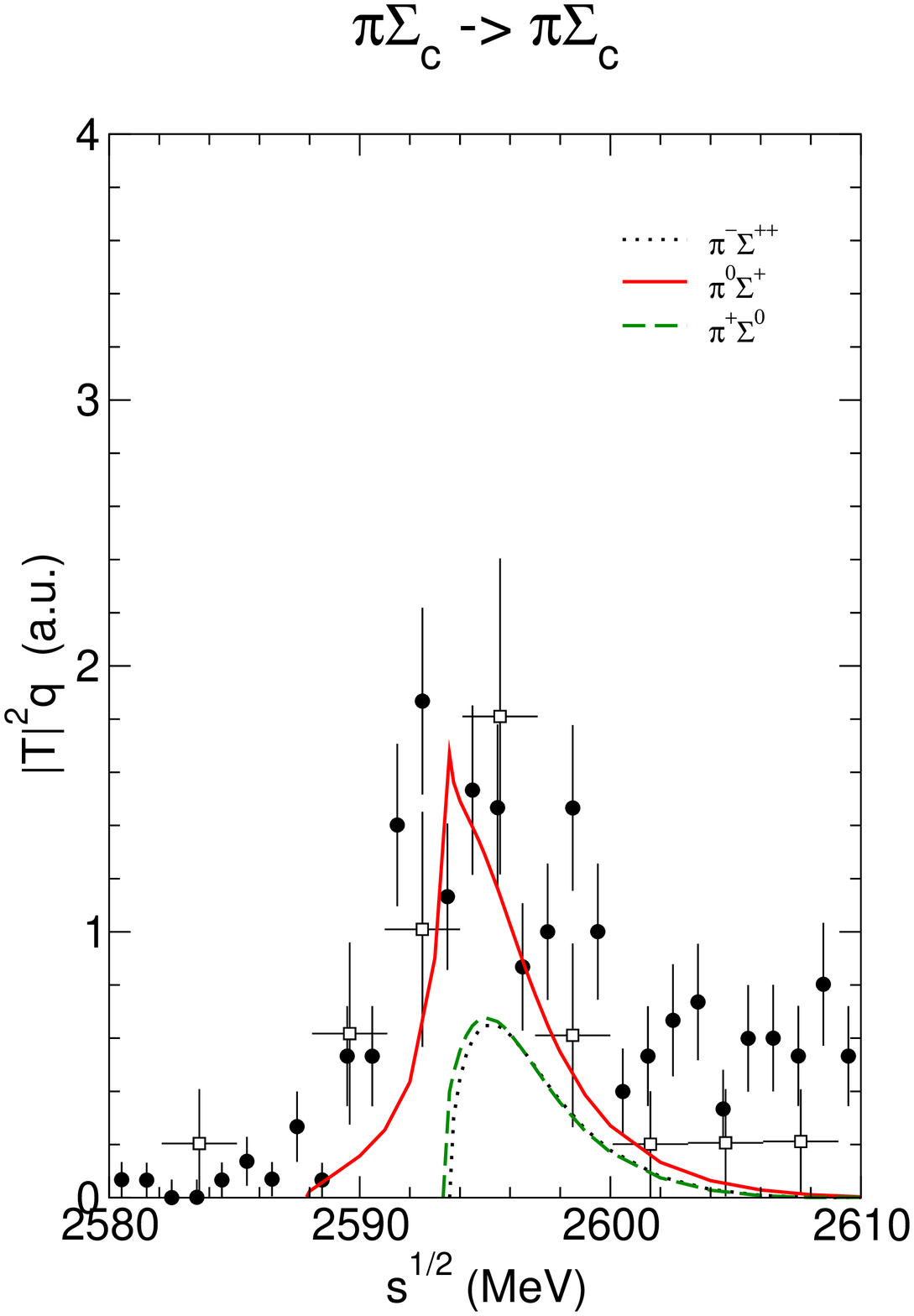}
\end{minipage}
\caption{Left and middle plots: $DN$ cross sections for the isospin
$I=0$  and $I=1$  channels as a function of 
$\epsilon = \sqrt{s} - m_N - m_{D}$. Results for the J\"ulich model \cite{haid11} (solid lines), the SU(4) WT model \cite{Mizutani06} (dashed lines) and SU(8) WT scheme \cite{GarciaRecio09} (dashed-double-dotted lines) are displayed. Right plot:
$\pi\Sigma_c$ invariant mass spectrum predicted by the J\"ulich
$DN$ meson-exchange model. We show also data for the 
$\pi^+\pi^-\Lambda^+_c$ invariant mass distribution 
taken from \cite{ARGUS2} (squares) and \cite{CLEO2} (circles).
}
\label{figdn}
%\end{center}
\end{figure}

The SU(4) extension of the J\"ulich $\bar KN$ model to the $DN$ interaction \cite{haid11} generates narrow states in the $S_{01}$ and $S_{11}$ partial waves which we identify with the experimentally 
observed $\Lambda_c(2595)$ and $\Sigma_c(2800)$ resonances, 
respectively. Not surprisingly, we find an additional pole in the 
$S_{01}$ partial wave,  located close to the other one with a larger width, similarly to the $\bar KN$ sector. Our model also generates a further state, name\-ly in 
the $P_{01}$ partial wave at 2804 MeV, i.e. just below the $DN$ threshold.  We are tempted to 
identify this state with the $\Lambda_c(2765)$ resonance,  whose quantum
numbers are not yet established~\cite{PDG}.

These results can be compared to previous works on the $DN$ interaction. The SU(4)  WT model of Ref.~\cite{Mizutani06} obtains three $S_{01}$ and two $S_{11}$ states up to 2900 MeV ~\cite{GarciaRecio09}.  Among them, there is a $S_{11}$ state at $2694 \ {\rm MeV}$ with a width of $\Gamma= 153 \ {\rm MeV}$ that strongly couples to the $DN$ channel, with 
similar effects as the $S_{11}$ resonance of our model. On the other hand, the SU(8) $DN$ WT model \cite{GarciaRecio09} predicts even more states in this energy region.

Those resonant states will have a determinant role in some $DN$ scattering observables close to the $DN$ threshold, such as scattering lengths and cross sections. Within the J\"ulich model we obtain the following scattering lengths for different isospin ($I$): $a_{I=0}$ $=$ $-$0.41 + $i$\,0.04  fm and $a_{I=1}$ $=$  $-$2.07 + $i$\,0.57 fm. The large value of the real part of $a_{I=1}$ is due to our  $S_{11}$ resonant state. The $S$-wave scattering lengths predicted by our model and by the SU(4) WT approach turn out to be very similar qualitatively for the $I=1$ as well as for the $I=0$ 
channel. This is in contrast to the SU(8) WT model of \cite{GarciaRecio09} which predicts radically different scattering lengths due to the different resonant structure.

The $DN$ cross sections for $I=0$ and $I=1$ are presented in the left and middle plots of Fig.~\ref{figdn}. We show results of the J\"ulich $DN$ model \cite{haid11} based on the parameter set that reproduces the positions of the $\Lambda_c(2595)$ 
and $\Sigma_c(2800)$ of the Particle Data Group (solid lines).  The $DN$ cross sections of the SU(4) WT model of Ref.~\cite{Mizutani06} (dashed lines) and of the 
SU(8) WT model of Ref.~\cite{GarciaRecio09} (dash-double-dotted lines) 
are also displayed. The $DN$ cross sections of the SU(4) WT approach of Ref.~\cite{Mizutani06}  show a similar behaviour as the one of the J\"ulich model for the $I=1$ channel.  As already discussed above, this model generates likewise poles in the $S_{11}$ 
partial wave. In case of the $I=0$ channel there are pronounced differences. But this is
not surprising because the SU(4) WT model yields only $S$-wave contributions
while the results of the J\"ulich model are dominated by the $P$-wave.
Overall larger differences are seen in comparison to the results for the
SU(8) WT model.

Finally, we would like to discuss the $\Lambda(2595)$ resonance
and the role of the near-by $\pi\Sigma_c$ threshold. The excited charmed baryon  $\Lambda_c$(2595) was first observed by the CLEO collaboration \cite{CLEO1} and  later confirmed by  E687 \cite{E6872} and ARGUS \cite{ARGUS2}, appearing
as a pronounced peak in the invariant mass distribution of the $\pi^+\pi^-\Lambda_c^+$ channel.

In Fig. \ref{figdn} (right plot) we present results for the $\pi\Sigma_c$ invariant mass spectrum
in the particle basis for $\pi \Sigma_c \rightarrow \pi \Sigma_c$ which illustrate the subtle effects of the slightly different thresholds of the $\pi^+\Sigma_c^0$, $\pi^0\Sigma_c^+$, 
and $\pi^-\Sigma_c^{++}$ channels on the various invariant mass 
distributions. Similar invariant mass distributions are obtained for $D N \rightarrow \pi \Sigma_c$ channel \cite{haid11}.

The $\pi^0\Sigma_c^+ \to \pi^0\Sigma_c^+$ channel resembles very much the measured signal 
and one can imagine that smearing out our results by the width of the $\Sigma_c^+$, which is roughly 4~MeV \cite{CLEO2,PDG}, would yield a fairly good fit to the data. However, experimentally it was found that the $\Lambda_c$(2595) decays predominantly into the $\pi^+\Sigma_c^0$ and $\pi^-\Sigma_c^{++}$ channels with a branching fraction in the range of 66\% \cite{ARGUS2} to close to 100\% \cite{CLEO2}. Smearing out the corresponding results with the significantly
smaller and better known widths of the $\Sigma_c^0$ and $\Sigma_c^{++}$,
of just 2~MeV \cite{PDG}, would still leave many of the events found below the nominal $\pi^+\Sigma_c^0$ and $\pi^-\Sigma_c^{++}$ threshold unexplained, especially for the CLEO experiment \cite{CLEO2}. Of course, the presence of the $\pi\pi \Lambda_c^+$ channel should be incorporated in our model. But, in any case, it would be important to confirm the new 
CLEO data by independent measurements of the $\pi\pi\Lambda_c^+$
and $\pi\Sigma_c$ mass spectra in the region of the $\Lambda_c(2595)$.

\begin{theacknowledgments}
This work is partially supported by the Helmholtz Association through funds
provided to the virtual institute ``Spin and strong QCD'' (VH-VI-231), by
the EU Integrated Infrastructure Initiative  HadronPhysics2 Project (WP4
QCDnet), by BMBF (06BN9006), by DFG 
(SFB/TR 16, ``Subnuclear Structure of Matter'') and FPA2010-16963 from DGI funds. 
G.K. acknowledges
financial support from CAPES, CNPq and FAPESP (Brazilian agencies). 
L.T. acknowledges support from the Ramon y
Cajal Research Programme (Ministerio de Ciencia e Innovaci\'on)
 \end{theacknowledgments}



\bibliographystyle{aipproc}   % if natbib is available

\end{document}